\newcommand{\Lm}{L_M}
\newcommand{\Am}{A_M}
\newcommand{\Bm}{B_M}
\newcommand{\Cm}{C_M}
\newcommand{\dm}{d_M}
\newcommand{\gs}{g_\textrm{S}}
\newcommand{\vmustark}{\vec{\mu}_M}
\newcommand{\mustark}{\mu_M}
\newcommand{\pstark}{\bar{\bar{\Theta}}_M}
\newcommand{\finect}{\alpha}
\newcommand{\enul}{\epsilon_0}
\newcommand{\edc}{\epsilon}
\newcommand{\nh}{n_h}
\newcommand{\gnul}{\Gamma_0}
\newcommand{\Er}{\mathbf{E}(\mathbf{r})}

\documentclass[prb,reprint,showpacs]{revtex4-1}
\usepackage{graphicx,epsfig,hyperref}
\usepackage{amsmath}
\usepackage{amssymb}
\usepackage{amsthm}

\begin{document}

%Title of paper
\title{Optical tracing of multiple charges in single-electron devices}

\author{Sanli Faez}
\email[]{Email: Faez@physics.leidenuniv.nl}
\affiliation{Huygens-Kamerlingh Onnes Laboratorium, Universiteit Leiden, Postbus 9504, 2300 RA Leiden, The Netherlands}
\author{Sense Jan van der Molen}
\affiliation{Huygens-Kamerlingh Onnes Laboratorium, Universiteit Leiden, Postbus 9504, 2300 RA Leiden, The Netherlands}
\author{Michel Orrit}
\affiliation{Huygens-Kamerlingh Onnes Laboratorium, Universiteit Leiden, Postbus 9504, 2300 RA Leiden, The Netherlands}

\date{\today}

\begin{abstract}
Single molecules that exhibit narrow optical transitions at cryogenic temperatures can be used as local electric-field sensors. We derive the single charge sensitivity of aromatic organic dye molecules, based on first principles. Through numerical modeling, we demonstrate that by using currently available technologies it is possible to optically detect charging events in a granular network with a sensitivity better than $10^{-5}e/\sqrt{\textrm{Hz}}$ and track positions of multiple electrons, simultaneously, with nanometer spatial resolution. Our results pave the way for minimally-invasive optical inspection of electronic and spintronic nanodevices and building hybrid optoelectronic interfaces that function at both single-photon and single-electron levels.
\end{abstract}

% insert suggested PACS numbers in braces on next line
% 42.25.Dd    Wave propagation in random media


%\maketitle must follow title, authors, abstract, \pacs, and \keywords
\maketitle

%\section{Introduction:}
Sixty years after the invention of transistors, conduction electrons in solids are still the workhorses of information processing. Nanotechnology has enabled shrinking the size of transistors to a level where charge granularity and quantum effects emerge. Meanwhile, optical communication has widely replaced electronic communication because of its larger bandwidth and lower losses. With these developments comes the demand for an optoelectronic interface that operates at both single-electron and single-photon limits.

Single organic molecules in the solid state were first detected more than two decades ago by Moerner and Kador~\cite{moerner_optical_1989}, using absorption spectroscopy, and by Orrit and Bernard~\cite{orrit_single_1990} based on fluorescence emission. Since then, single-molecule fluorescence microscopy techniques have found a vast variety of applications in physics, chemistry, and life sciences. More specifically, in the past decade, there has been a surge in their applications in quantum optics. For example, organic molecules have proven to be ideal candidates for single-photon sources~\cite{brunel_triggered_1999}. Magnetic resonance of a single electron spin has been demonstrated using spectroscopy on a single pentacene~\cite{kohler_magnetic_1993,wrachtrup_optically_1993}. Other aromatic molecules have been used for quantum state preparation and readout~\cite{gerhardt_coherent_2009}, even at room temperature~\cite{hildner_femtosecond_2011}. More recently, optical transistors were developed using the different energy levels of a single molecule~\cite{hwang_single-molecule_2009} and coherent coupling between distant molecules has been achieved~\cite{rezus_single-photon_2012}.

Empowered by these accomplishments, we can now consider single molecules as building blocks for integrated photonics circuits and hybrid quantum devices~\cite{wallquist_hybrid_2009}. Successful insertion of these molecules into nanochannels, without any degradation of their coherence properties, has already allowed for demonstration of coherent interaction of light with several molecules in a dielectric waveguide~\cite{faez_coherent_2014}. This result has also paved the way for on-demand insertion of organic molecules in other electronic nanodevices as quantum nanoprobes.

In this article we demonstrate that a single organic molecule can be used for optical detection and locating individual electrons. At cryogenic conditions the optical transition linewidth of some aromatic molecules in solid host matrices is narrower than 30~MHz~\cite{tamarat_ten_2000}. We illustrate how the movement of a single electron in the micrometer surrounding of such a molecule would change its transition frequency by several times its linewidth. As a result, the electron can be optically traced by following the molecular lineshift. To demonstrate the speed and sensitivity of this sensing method, we consider a generic single-electron device that consists of a single metallic island and two electrodes. Further, we present our simulation results on feasibility of locating the position of several electrons with nanometer spatial resolution. These simulations are implemented considering a nanoparticle chain that exhibits Coulomb blockade as the simplest relevant example. All suggested measurements can be performed based on currently available knowledge and technology.

%\subsection{\label{sec:Molprobes}Molecular nanoprobes}
At moderately low temperatures, around 2 K, certain organic molecules present a very strong zero-phonon line (ZPL) on their electronic transition at optical frequencies (300 THz) with a lifetime-limited linewidth of less than 30 MHz~\cite{tamarat_ten_2000}. The nominal quality factor corresponding to this resonance is better than $10^7$. As such, each molecule can be seen as a highly sensitive local probe of its environment. Experiments have shown that some of these molecules can acquire a dipole moment difference as large as 1~Debye ($3.3\times10^{-30}~\textrm{C}\cdot\textrm{m}$) between their electronic ground and excited states~\cite{orrit_stark_1992}. The transition frequency is thus very sensitive to the local electric field. For example, a sizable fraction of terrylene molecules in a p-terphenyl crystal exhibit a linear lineshift of more than 3~MHz/(kV/m) in response to external electric field~\cite{kulzer_nonphotochemical_1999}, despite the centrosymmetry of the terrylene molecule. This anomalously high linear response is induced by the deformation of the molecular orbital in the crystal~\cite{bordat_anomalous_2000}. The electrostatic field of a single electron at a 100~nm distance is roughly 150 kV/m. This field is high enough to shift the transition frequency of these molecules by more than three times its 42~MHz linewidth~\cite{kummer_terrylene_1994}. Considering this gigantic sensitivity and the small size of these molecules, Caruge and Orrit have suggested to detect electronic currents in semiconductors using organic molecules as nanoprobes~\cite{caruge_probing_2001}. Later, Plakhotnik has shown that subnanometer displacement of a single electron in a dielectric medium can be detected by simultaneously looking at the lineshifts of multiple molecular probes~\cite{plakhotnik_single-molecule_2006, plakhotnik_sensing_2007}. Here, we will use first principles and a simple electron-in-a-box model to identify the origin of this sensitivity and routes to its optimization.

The common feature in the conjugated molecules that show lifetime-limited linewidth is their rigid backbone. Quantum mechanical calculations based on box boundary conditions have been relatively successful in predicting the optical transition frequencies of this type of molecules\cite{platt_box_1954}. For our discussion, considering the simplest model of an electron in a one-dimensional box is sufficient. The eigenfunctions of this model correspond to molecular $\pi$-orbitals that are filled with a total number or $N$ electrons. The transition frequency from the highest occupied molecular orbital (HOMO) to the lowest unoccupied molecular orbital (LUMO) is given by
\begin{eqnarray}\label{eq:omegatransition}
\nu&=&\frac{(N+1)h}{8 m_e \Lm^2},
\end{eqnarray}
where $h$ is the Planck constant, $m_e$ is the electron mass, and $\Lm\equiv \Am a_0$ is the symbolic length of the box. We take the Bohr radius $a_0=52.9(2) pm$ as the smallest relevant length scale for defining molecular dimensions. The numerical factor $\Am$ can be chosen such that the values of the optical transition in the model match the experimental result. It helps our discussion to reformulate Eq.~(\ref{eq:omegatransition}) as
\begin{eqnarray}\label{eq:omegareform}
\Lm\nu&=&\frac{\pi (N+1)c\finect}{4 \Am},
\end{eqnarray}
with $c$ the speed of light in vacuum and $\finect \equiv \frac{h}{2\pi m_e c a_0}$ the fine structure constant. Next, we look at the frequency width of the transition given by the Fermi golden rule:
\begin{eqnarray}\label{eq:gammafermi}
\gnul&=&\frac{8 \pi^2 \nh \dm^2 \nu^3}{3 \enul h c^3}.
\end{eqnarray}
The refractive index of the medium at frequency $\nu$ and the vacuum permittivity are denoted by $\nh$ and $\enul$, and $\dm$ is the transition dipole moment. This dipole moment can be written as
\begin{eqnarray}\label{eq:defdm}
\dm&=&\Bm e \Lm,
\end{eqnarray}
where $e$ is the elementary charge and $\Bm$ is a system-dependent numerical prefactor, which can be calculated based on either first principles or experimental results. Dividing Eq.~(\ref{eq:omegareform}) by Eq.~(\ref{eq:gammafermi}) yields the nominal quality factor of the molecular transition
\begin{eqnarray}\label{eq:qfactor}
Q\equiv\frac{\nu}{\gnul}&=&\frac{3\Am^2}{\pi^4(N+1)^2\nh\Bm^2}\left(\frac{1}{\finect}\right)^3.
\end{eqnarray}
The system dependent prefactor on the right hand side of Eq.~(\ref{eq:qfactor}) happens to be in the order of unity for the aromatic molecules that are relevant to the subject of this paper. The narrow linewidth of the molecular transitions, with high quality factors of $Q=10^7$, have a well-known connection with the small value of the fine structure constant~\cite{cohen-tannoudji_atom-photon_1998, devoret_circuit-qed:_2007}.

We proceed with the calculation of the lineshift in response to external electric field. In general, this Stark shift can be written as
\begin{eqnarray}\label{eq:starkshift}
h\gs&=&- \vmustark \cdot \Er - \frac{1}{2} \Er \cdot \pstark \cdot \Er,
\end{eqnarray}
where $\vmustark$ and $\pstark$ are the changes in the molecular dipole moment and molecular polarizability tensor upon excitation. $\Er$ is the local (quasi)static electric field at the position of the molecule. The local field is mainly determined by the immediate surrounding of the molecule. In common single molecule spectroscopy measurement, the environment configuration is frozen, hence the changes in the transition energy can still be related directly to the external field provided that the local field effects are properly included~\cite{vallee_microscopic_2005}. Implementing this correction does not influence the present discussion. For the purpose of electric field sensing, we are interested in large Stark shifts, which are mainly found for emitters with broken centrosymmetry, either internally or induced by the host matrix. In this case, the linear term in Eq.~(\ref{eq:starkshift}) dominates. Similar to the transition dipole moment, we relate the linear dipole change to the elementary charge and the nominal length of the molecule by
\begin{eqnarray}\label{eq:defmu}
\mustark \equiv |\vmustark|&=&\Cm e \Lm.
\end{eqnarray}
The change of electric field due to small radial displacement $\Delta r$ of a single electron charge at a distance $r$ from the molecule is given by
\begin{eqnarray}\label{eq:dEr}
\Delta E(r)&=&\frac{1}{2\pi\edc\enul}\left(\frac{\Delta r}{r^3}\right) ,
\end{eqnarray}
with $\edc$ the static dielectric constant. To define a figure of merit for sensitivity, the Stark shift due to this field variation should be compared with the natural linewidth. By inserting Eq.~(\ref{eq:dEr}) in Eq.~(\ref{eq:starkshift}) and dividing it by Eq.~(\ref{eq:gammafermi}) and considering the emission wavelength in vacuum $\lambda=c/\nu$ we obtain
\begin{eqnarray}\label{eq:shiftratio}
\frac{\gs}{\gnul}&=&\frac{3 \Cm}{16 \pi^3 \edc \nh \Bm^2}\left(\frac{\lambda}{r}\right)^3\frac{|\Delta r|}{\Lm}.
\end{eqnarray}
This equation is the key result of this discussion. Apart from the numerical prefactor, this general argument holds for any single electron two-level emitter that decays solely by spontaneous emission and is protected from other sources of decoherence like for example nitrogen vacancy centers in diamond~\cite{dolde_electric-field_2011}. Although modern spectroscopy techniques enable us to measure lineshifts much smaller than a linewidth, we choose to be conservative and call a nanoprobe sensitive if $\gs>\gnul$ (strong coupling). Equation~(\ref{eq:shiftratio}) shows that such an emitter is highly sensitive to the movement of a single elementary charge within its optical near-field volume. This volume, set by the radiation wavelength, is generally much larger than the emitter size. The same interpretation, for example, clarifies the extreme sensitivity of a superconductive Josephson qubit, with a microwave transition, to charge fluctuations in its extended environment~\cite{bouchiat_quantum_1998}. In table~\ref{tbl:ABC}, we list the experimentally measured properties of several aromatic molecules that exhibit, under cryogenic conditions, both a lifetime-limited ZPL and a large linear response to external electric field.

%\section{Single charge detection}
The high sensitivity of molecular nanoprobes to the presence of charges in their extended surrounding allows for tracing conduction electrons in solid-state systems. In the following, we use numerical modeling to demonstrate the working principles of this method, its time resolution, and its spatial accuracy. For this purpose, we consider the most generic geometries that are commonly used for modeling single-electron studies. We assume a planar device made of flat conducting electrodes and metallic islands in between them. The organic host crystal is doped with the aromatic molecules and put on top of the devices. This layer is separated by a thin insulating layer that covers the electronic device to eliminate possibility of charge transfer between the device and the crystal. As molecular nanoprobe we take the system of terrylene in p-terphenyel due to its superior brightness and spectral stability. The zero-phonon transition emits at $\lambda=579$~nm with a linewidth of $\gnul=42$~MHz and an experimentally measured Stark coefficient of up to 3~MHz/(kV/m)~\cite{kulzer_nonphotochemical_1999}.

\begin{figure}[t]
\includegraphics[width=8cm]{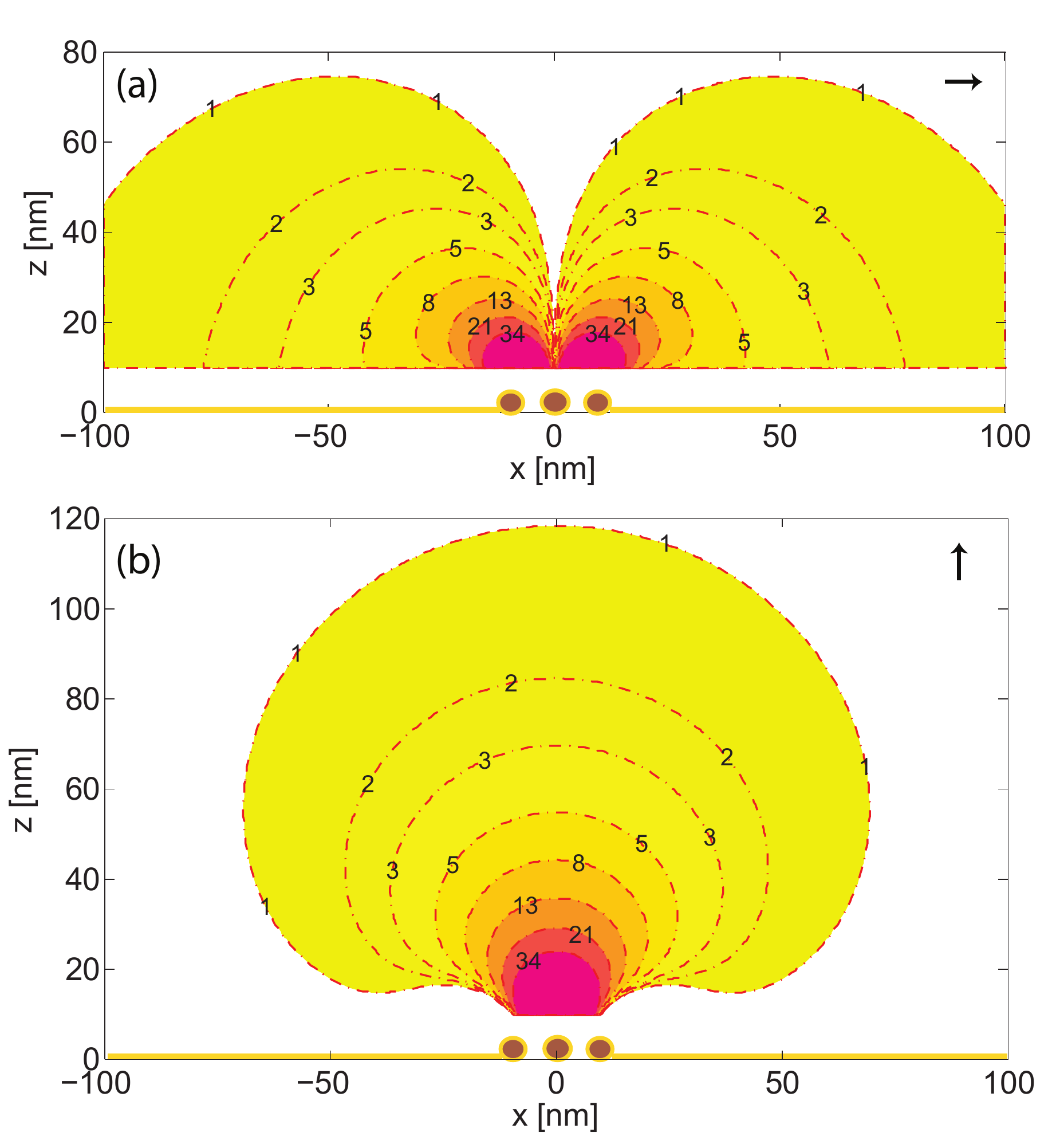}
\caption{\label{fig:singleisland} Contour plot of the coupling constant ($|\gs/\gnul|$) as a function position above a single-charged island between grounded electrodes for (a) parallel and (b) perpendicular dipole change orientations of the molecules with respect to the substrate. Parameters: $\gnul=42$~MHz, $\mustark/e=6$~pm, $\edc=2.5$, $C_g/C_j=3$.}
\end{figure}

%\subsection{Single electron box}
The simplest electronic device that exhibits single charge transport is the electron box. It consists of a nanoscale metallic island capacitively coupled to junction electrodes. We consider first the regime of large tunneling resistivity so that the charging and discharging times $t=RC$ are larger than the required measurement time of about a few milliseconds. Here, $R$ and $C$ are the total resistance and the total capacitance between the island and the rest of the system. To show the sensitivity as a function of molecule-island separation, we consider a single charge on the island while the electrodes are connected to ground. This will allow us to investigate the effect of field screening by the junctions. For simplicity, we model them as small grounded spheres. The capacitance between the island and each junction is denoted by $C_j$ and the gate capacitance is $C_g$. Considering this simple geometry, the position dependence of molecule-electron coupling constant $\gs/\gnul$, i.e. the lineshifts of single terrylene molecules after adding one electron to the island, are calculated and the results are depicted in Fig.~\ref{fig:singleisland}(a) and (b) for dipole change orientation parallel and perpendicular to the substrate, respectively.
This plot demonstrates the superior charge sensitivity of a single organic molecule. We find that the transition of a molecule located 10~nm away from the island reacts to a charge difference of $10^{-2}e$ by a full linewidth shift, hence is no longer excited at the initial excitation frequency. The total emission rate of the molecule is limited by its upper-state lifetime of a few nanoseconds. In practice, due to the limited collection efficiency, it is possible to detect around $10^6$ photons per second on a single photon detector without any significant line-broadening~\cite{wrigge_efficient_2008}. With these figures, the charge sensitivity of such an optical detection can easily reach $10^{-5}e/\sqrt{\textrm{Hz}}$, which is on par with the highly celebrated sensitivity of single electron transistors~\cite{devoret_amplifying_2000}. This is just an empirical value for the simplest configuration presented here. In fact, with a proper design, the electrodes will act as a plasmonic antenna that enhances the emission rate by at least ten times~\cite{kinkhabwala_large_2009}.

%\subsection{Positioning accuracy}
Next to sensing the presence of charges on a metallic island, molecular nanoprobes can also be used to localize the position of electrons in a granular network. In cryogenic single molecule spectroscopy, molecules are distinguished through their ultra-narrow spectral response and hence they can be localized with an accuracy far beyond the diffraction limit~\cite{betzig_proposed_1995}. Unlike room-temperature localization microscopy, cryogenic spectroscopy measurements can handle molecular concentrations as high as $10^4$ per cubic micrometer. Furthermore, given the unlimited photon budget, the position of each molecule can be determined with an accuracy better than one nanometer~\cite{hettich_nanometer_2002,weisenburger_cryogenic_2014}.
\begin{figure}[h]
\includegraphics[width=8.5cm]{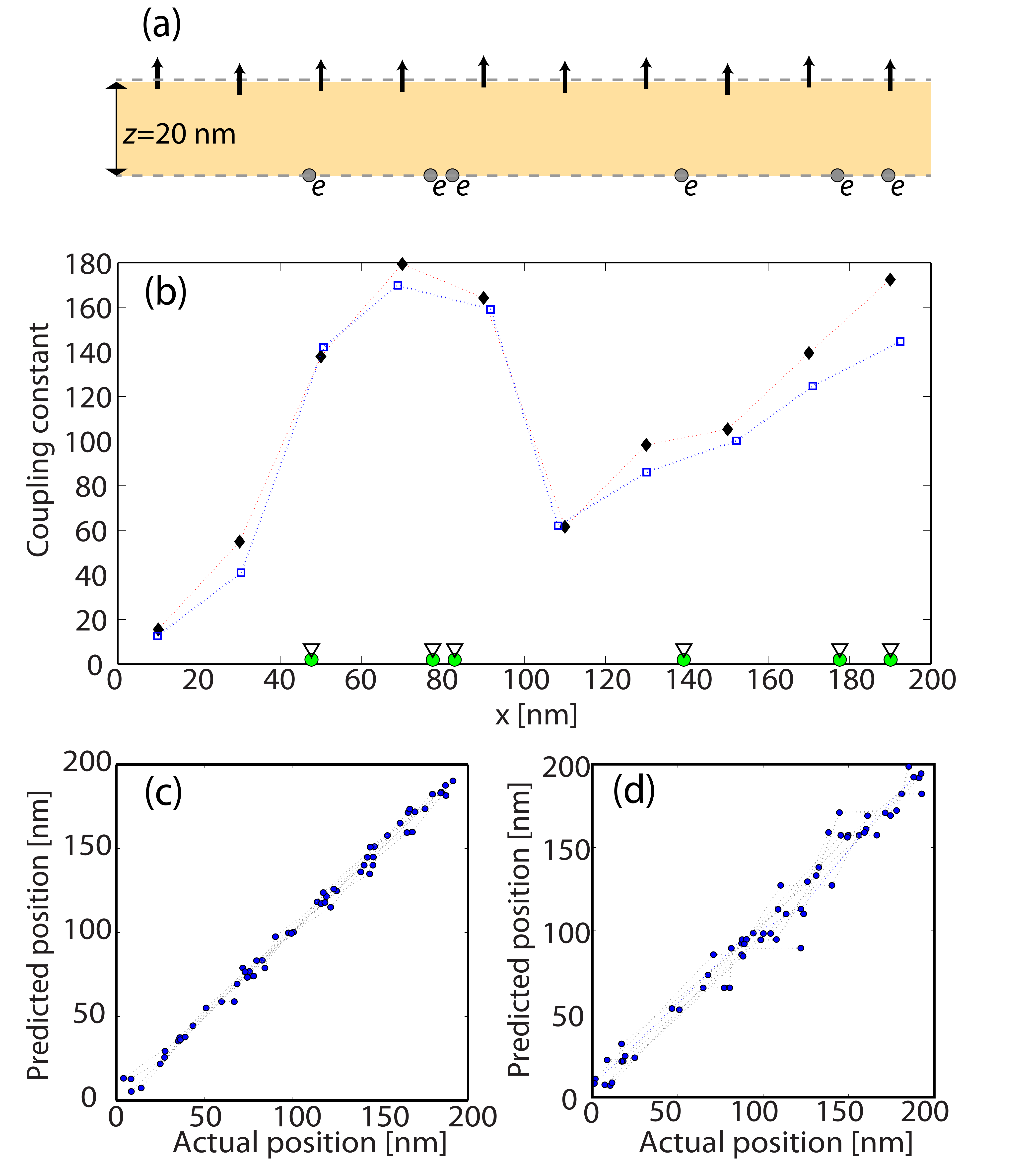}
\caption{\label{fig:network} (a) The modeling procedure: Probe molecules, depicted by arrows, are evenly separated at an average spacing of 20~nm from each other on the $z=20$~nm line. All molecules are aligned and respond maximally to electric field in the $z$ direction. Charges are randomly distributed on the $z=0$ line. The positions for a typical realization with $n=6$ electrons are indicated by circles (b) The corresponding lineshifts relative to the linewidth of the molecules, induced by the presence of electrons, are indicated by squares. The values for the lineshifts are changed randomly by $\pm3$ percent of the actual values and the molecule coordinates are changed by adding a random displacement in both $x$ and $z$ directions. The new values for this specific realization are indicated by filled diamonds, from which the charge distribution is reconstructed (downward arrows close to the lower axis). Parameters: 10 molecules, 6 electrons, $\gnul=42$~MHz, $\mustark/e=6$~pm, $\edc=2.5$. (c) The predicted positions of 6 electrons in 10 random configuration compared with the actual positions considering a localization inaccuracy of $\pm1$~nm in the $x$ direction and $\pm3$~nm in the $z$ direction. The 6 datapoints corresponding to each realization are connected by dotted lines for clarity. (d) Similar to (b) but considering an exaggerated localization inaccuracy of $\pm10$~nm in both $x$ and $z$ directions.}
\end{figure}

To determine the coordinates of a single electron in three dimensions, the simultaneous lineshift of three or more molecules can be used. This triangulation method and its accuracy have been previously discussed in the context of studying charge transfer in a chemical bond~\cite{plakhotnik_single-molecule_2006}. In that work an optimal localization precision of about 10~pm has been suggested based on Monte-Carlo simulations. Here, we present a more comprehensive analysis, considering the experimentally relevant parameters, for the goal of imaging distribution of multiple charges and their dynamics in electronic nanodevices.

%\subsection{\label{sec:chain}Nanoparticle chains}
%

To visualize parallel detection of several conduction electrons, we take a nanoparticle chain device at cryogenic temperatures as a generic example. This and other granular systems have been in use from the early days of measuring single charge transport~\cite{middleton_collective_1993}, but are still the topic of active research in the context of molecular electronics~\cite{gorter_possible_1951, dayen_enhancing_2013}. For clarity of presentation, we restrict our discussion to the case of a linear chain of capacitively coupled small islands. We note that the same procedure can be applied to two-dimensional network of particles and our conclusions are not restricted to a linear chain.
For identifying the position of $n$ electrons on such a linear chain, in the most general case, there are $n+1$ nanoprobes necessary. The main sources of inaccuracy are the localization error of the molecules and uncertainties in their Stark factors. To increase the accuracy, it is necessary to probe a larger number of molecules than electrons and thereafter solve an over-parameterized inverse problem. These molecules are distinguished both in position and in excitation frequency and following the spectral shifts of several molecules, simultaneously, is straightforward.

We then perform Monte-Carlo simulations with synthetic errors to estimate the propagating errors due the positioning inaccuracy. The modeling scheme is as follows. We consider 10 probe molecules evenly spaced on a line parallel to the substrate with a intermolecular separation of 20~nm. The spacing between the molecules and the substrate is also set to 20~nm as schematically presented in Fig.~\ref{fig:network}(a). A total number of $n<10$ electrons are randomly distributed inside a 200~nm interval on the substrate and the induced molecular lineshifts relative to the charge-neutral state are calculated. The experimental uncertainty is simulated by imposing $\pm1$~nm localization errors on the lateral position of the molecules, varying their separation from the substrate within $\pm3$~nm, and varying the magnitude of the shifts by $\pm3$ percent. One example is plotted in Fig.~\ref{fig:network}(b). Note that although these localization inaccuracies are much smaller than the optical diffraction limit, they are now routinely accessible in cryogenic conditions using single molecule localization techniques\cite{weisenburger_cryogenic_2014}. The position of electrons is recalculated based on a least-square optimization routine.  The mean deviation is estimated by taking the average of the absolute difference between the reconstructed and actual positions of the electrons for many realizations. As an example, reconstructed positions are plotted versus the actual positions for 10 realizations of $n=6$ electrons in Fig.~\ref{fig:network}(c). This plot, highlights the high accuracy of locating multiple electrons simultaneously based on our suggested technique. To emphasize on the robustness of this method, we have also followed the similar reconstruction routine considering an exaggerated localization inaccuracy of $\pm10$~nm in both $x$ and $z$ directions and the results are depicted in Fig.~\ref{fig:network}(d).

\begin{figure}[t]
\includegraphics[width=7cm]{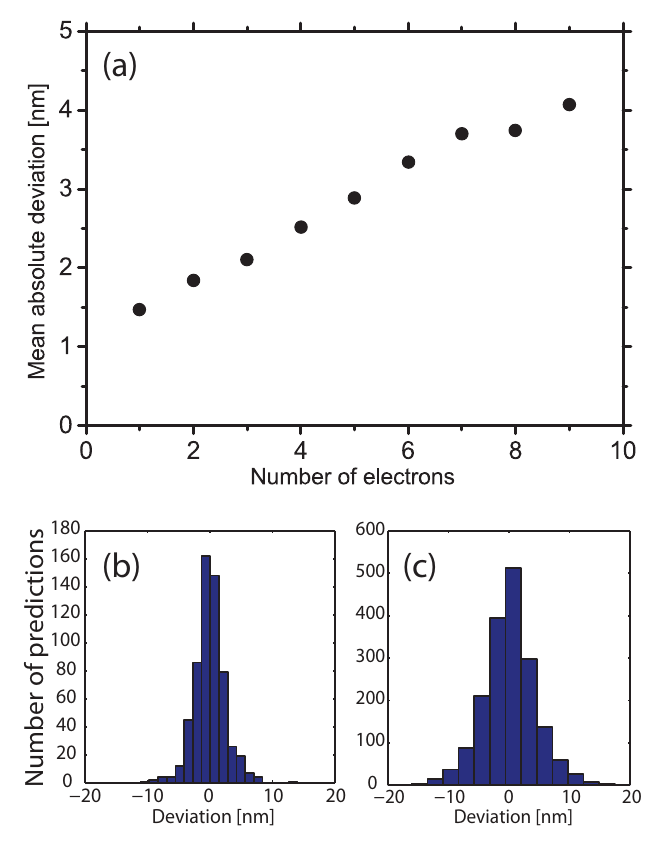}
\caption{\label{fig:delx} (a)The reconstruction inaccuracy for $n$ electrons when using 10 probe molecules. (b) The deviation histogram for $n=2$ based on 300 realizations. (c) Same for $n=6$. Simulations parameters are the same as in Fig.~\ref{fig:network}(b).}
\end{figure}

The simulation results are plotted in Fig.~\ref{fig:delx} as a function of $n$. For each point in the plot, 300 realization are simulated. These results show a localization accuracy of 1.5~nm when a single electron is probed using 10 molecules. Most notably, this value is even smaller than the optical localization inaccuracy that we considered for each individual molecule. The mean deviation between predicted and actual positions increases almost linearly with the number of electrons that are simultaneously probed, while keeping the number of probe molecules constant. This result proves the scalability of our suggested technique for parallel tracing of multiple electrons.
%%\subsection{Future directions}
Although electronic amplifiers allow routine measurement of minuscule currents nowadays, several of these techniques fail in the total Coulomb blockade regime where current is vanishingly small. Some scanning methods such as Kelvin probe microscopy~\cite{sadewasser_kelvin_2011} also function based on static charge mapping. However, scanning methods like this are inherently slow and currently incapable of addressing multiple locations in parallel. In our technique, as demonstrated above, several molecules in the 500-nanometer surrounding of a device can act as simultaneous probes. This allows parallel mapping of charge distribution with an information density and speed that is unsustainable for scanning probe techniques.

We emphasize that our choice of a simple single-electron device was for demonstrating the working principles of this minimally-invasive optical tracing technique. As we have discussed, the main strengths of using molecular nanoprobes for tracing charges, in comparison with conventional electrometrical techniques, is in visualizing the slow dynamics of multiple static charges. However, this method can also be extended, by fast and shot-noise limited photon detection, to measure the full counting statistics of charge transport~\cite{levitov_electron_1996}, which is a more general form of current measurement. The mean current can then be determined from the peak of the charge fluctuations power spectrum~\cite{bylander_current_2005}. This is somehow similar to the correlation measurements that are commonly used to quantify for example the triplet-state lifetime in molecules or other sources of spectral diffusion~\cite{nicolet_intermolecular_2006}. Here, the fluorescence signal from a molecule is measured by time-correlated photon counting, while the frequency of the excitation laser is fixed. Given the brightness of these molecules, this time trace can be recorded at 10 MHz rate with a time stamp accuracy of 100 picoseconds. In the charging interval between successive tunneling events, the molecule line is shifted and hence it will be dark. The rate of charging events and its duration can be recovered from the auto-correlation of the fluorescence time trace.

\begin{table*}[t]	
\caption{\label{tbl:ABC} Summary of optical transition properties for dibenzoterylene (DBT), dibenzoanthranthene (DBATT), terrylene (Ter) in two host matrices, as well as CdSe colloidal nanocrystals for comparison. }
\begin{ruledtabular}
\begin{tabular}{lccccl}
    Emitter/Host & $N$ & $\lambda$ [nm] & $\gnul$ [MHz] & $\mustark / e$ [pm]\footnote{Typical value at the waist of the distribution} & Ref. \\
    \hline
    DBT/anthracene & 38 & 784 & 30 & 1.2 & \onlinecite{nicolet_single_2007}\\
    DBATT/hexadecane & 30 & 589 & 17 & 6 & \onlinecite{brunel_stark_1999}\\
    Ter/p-terphenyl & 30 & 579 & 42 & 6 & \onlinecite{kulzer_nonphotochemical_1999}\\
    Ter/polyethylene & 30 & 575 & 42 & 10 & \onlinecite{orrit_stark_1992}\\
    CdSe/- & - & 620 & $10^6$\footnote{Experimental value, broadened by spectral diffusion} & 100 & \onlinecite{colvin_electric_1994, empedocles_quantum-confined_1997}\\

\end{tabular}
\end{ruledtabular}
\end{table*}

In conclusion, we have presented a new method for measuring static charge distribution in electronic devices such as single electron transistors or granular networks and have demonstrated its sensitivity. Our approach is an alternative to scanning probe techniques, such as Kelvin probe microscopy, for investigation of charge distribution in solid state nanodevices. Using numerical simulations, we have illustrated a spatial accuracy of 1 nm for positioning a single electron and less than 3 nm for tracking of up to 6 electrons simultaneously, using 10 probe molecules. The required organic substances are soft and easy-to-handle materials, and can be placed on top of many other conducting or semiconductor devices with minimum difficulty. Since our electron sensing method functions at single-photon, single-molecule, and single-electron levels, it sets the basis for building hybrid quantum devices.

This research was supported by the Netherlands Organization for Scientific
Research, NWO (Vidi grant SJvdM), and European Research Council (Advanced Grant SiMoSoMa).

\bibliographystyle{ieeetr}
\bibliography{eprobe}

\begin{thebibliography}{10}

\bibitem{moerner_optical_1989}
W.~E. Moerner and L.~Kador, ``Optical detection and spectroscopy of single
  molecules in a solid,'' {\em Physical Review Letters}, vol.~62,
  pp.~2535--2538, May 1989.

\bibitem{orrit_single_1990}
M.~Orrit and J.~Bernard, ``Single pentacene molecules detected by fluorescence
  excitation in a p-terphenyl crystal,'' {\em Physical Review Letters},
  vol.~65, pp.~2716--2719, Nov. 1990.

\bibitem{brunel_triggered_1999}
C.~Brunel, B.~Lounis, P.~Tamarat, and M.~Orrit, ``Triggered source of single
  photons based on controlled single molecule fluorescence,'' {\em Physical
  Review Letters}, vol.~83, pp.~2722--2725, Oct. 1999.

\bibitem{kohler_magnetic_1993}
J.~K\"{o}hler, J.~a. J.~M. Disselhorst, M.~C. J.~M. Donckers, E.~J.~J. Groenen,
  J.~Schmidt, and W.~E. Moerner, ``Magnetic resonance of a single molecular
  spin,'' {\em Nature}, vol.~363, pp.~242--244, May 1993.

\bibitem{wrachtrup_optically_1993}
J.~Wrachtrup, C.~von Borczyskowski, J.~Bernard, M.~Orrit, and R.~Brown,
  ``Optically detected spin coherence of single molecules,'' {\em Physical
  Review Letters}, vol.~71, pp.~3565--3568, Nov. 1993.

\bibitem{gerhardt_coherent_2009}
I.~Gerhardt, G.~Wrigge, G.~Zumofen, J.~Hwang, A.~Renn, and V.~Sandoghdar,
  ``Coherent state preparation and observation of rabi oscillations in a single
  molecule,'' {\em Physical Review A}, vol.~79, Jan. 2009.

\bibitem{hildner_femtosecond_2011}
R.~Hildner, D.~Brinks, and N.~F.~v. Hulst, ``Femtosecond coherence and quantum
  control of single molecules at room temperature,'' {\em Nature Physics},
  vol.~7, no.~2, pp.~172--177, 2011.

\bibitem{hwang_single-molecule_2009}
J.~Hwang, M.~Pototschnig, R.~Lettow, G.~Zumofen, A.~Renn, S.~G\"{o}tzinger, and
  V.~Sandoghdar, ``A single-molecule optical transistor,'' {\em Nature},
  vol.~460, pp.~76--80, July 2009.

\bibitem{rezus_single-photon_2012}
Y.~L.~A. Rezus, S.~G. Walt, R.~Lettow, A.~Renn, G.~Zumofen, S.~G\"{o}tzinger,
  and V.~Sandoghdar, ``Single-photon spectroscopy of a single molecule,'' {\em
  Physical Review Letters}, vol.~108, p.~093601, Feb. 2012.

\bibitem{wallquist_hybrid_2009}
M.~Wallquist, K.~Hammerer, P.~Rabl, M.~Lukin, and P.~Zoller, ``Hybrid quantum
  devices and quantum engineering,'' {\em Physica Scripta}, vol.~2009,
  p.~014001, Dec. 2009.

\bibitem{faez_coherent_2014}
S.~Faez, P.~T\"{u}rschmann, H.~R. Haakh, S.~G\"{o}tzinger, and V.~Sandoghdar,
  ``Coherent interaction of light and single molecules in a dielectric
  nanoguide,'' {\em {arXiv}:1407.2846}, July 2014.

\bibitem{tamarat_ten_2000}
P.~Tamarat, A.~Maali, B.~Lounis, and M.~Orrit, ``Ten years of single-molecule
  spectroscopy,'' {\em The Journal of Physical Chemistry A}, vol.~104,
  pp.~1--16, Jan. 2000.

\bibitem{orrit_stark_1992}
M.~Orrit, J.~Bernard, A.~Zumbusch, and R.~I. Personov, ``Stark effect on single
  molecules in a polymer matrix,'' {\em Chemical Physics Letters}, vol.~196,
  pp.~595--600, Aug. 1992.

\bibitem{kulzer_nonphotochemical_1999}
F.~Kulzer, R.~Matzke, C.~Br\"{a}uchle, and T.~Basch\'{e}, ``Nonphotochemical
  hole burning investigated at the single-molecule level:  stark effect
  measurements on the original and photoproduct state,'' {\em The Journal of
  Physical Chemistry A}, vol.~103, pp.~2408--2411, Apr. 1999.

\bibitem{bordat_anomalous_2000}
P.~Bordat, M.~Orrit, R.~Brown, and A.~W\"{u}rger, ``The anomalous stark effect
  of single terrylene molecules in p-terphenyl crystals,'' {\em Chemical
  Physics}, vol.~258, pp.~63--72, Aug. 2000.

\bibitem{kummer_terrylene_1994}
S.~Kummer, T.~Basch\'{e}, and C.~Br\"{a}uchle, ``Terrylene in p-terphenyl: a
  novel single crystalline system for single molecule spectroscopy at low
  temperatures,'' {\em Chemical Physics Letters}, vol.~229, pp.~309--316, Oct.
  1994.

\bibitem{caruge_probing_2001}
J.-M. Caruge and M.~Orrit, ``Probing local currents in semiconductors with
  single molecules,'' {\em Physical Review B}, vol.~64, p.~205202, Oct. 2001.

\bibitem{plakhotnik_single-molecule_2006}
T.~Plakhotnik, ``Single-molecule dynamic triangulation,'' {\em {ChemPhysChem}},
  vol.~7, pp.~1699--1704, Aug. 2006.

\bibitem{plakhotnik_sensing_2007}
T.~Plakhotnik, ``Sensing single electrons with single molecules,'' {\em Journal
  of Luminescence}, vol.~127, pp.~235--238, Nov. 2007.

\bibitem{platt_box_1954}
J.~R. Platt, ``The box model and electron densities in conjugated systems,''
  {\em The Journal of Chemical Physics}, vol.~22, pp.~1448--1455, Aug. 1954.

\bibitem{cohen-tannoudji_atom-photon_1998}
C.~Cohen-Tannoudji, J.~Dupont-Roc, and G.~Grynberg, {\em Atom-Photon
  Interactions: Basic Processes and Applications}.
\newblock Wiley, Mar. 1998.

\bibitem{devoret_circuit-qed:_2007}
M.~Devoret, S.~Girvin, and R.~Schoelkopf, ``Circuit-{QED}: How strong can the
  coupling between a josephson junction atom and a transmission line resonator
  be?,'' {\em Annalen der Physik}, vol.~16, pp.~767--779, Oct. 2007.

\bibitem{vallee_microscopic_2005}
R.~A.~L. Vall\'{e}e, M.~Van Der~Auweraer, F.~C. De~Schryver, D.~Beljonne, and
  M.~Orrit, ``A microscopic model for the fluctuations of local field and
  spontaneous emission of single molecules in disordered media,'' {\em
  {ChemPhysChem}}, vol.~6, pp.~81--91, Jan. 2005.

\bibitem{dolde_electric-field_2011}
F.~Dolde, H.~Fedder, M.~W. Doherty, T.~N\"{o}bauer, F.~Rempp,
  G.~Balasubramanian, T.~Wolf, F.~Reinhard, L.~C.~L. Hollenberg, F.~Jelezko,
  and J.~Wrachtrup, ``Electric-field sensing using single diamond spins,'' {\em
  Nature Physics}, vol.~7, pp.~459--463, June 2011.

\bibitem{bouchiat_quantum_1998}
V.~Bouchiat, D.~Vion, P.~Joyez, D.~Esteve, and M.~H. Devoret, ``Quantum
  coherence with a single cooper pair,'' {\em Physica Scripta}, vol.~1998,
  p.~165, Jan. 1998.

\bibitem{nicolet_single_2007}
A.~A.~L. Nicolet, P.~Bordat, C.~Hofmann, M.~A. Kol'chenko, B.~Kozankiewicz,
  R.~Brown, and M.~Orrit, ``Single dibenzoterrylene molecules in an anthracene
  crystal: Main insertion sites,'' {\em {ChemPhysChem}}, vol.~8,
  pp.~1929--1936, Sept. 2007.

\bibitem{brunel_stark_1999}
C.~Brunel, P.~Tamarat, B.~Lounis, J.~C. Woehl, and M.~Orrit, ``Stark effect on
  single molecules of dibenzanthanthrene in a naphthalene crystal and in a
  n-hexadecane shpol'skii matrix,'' {\em The Journal of Physical Chemistry A},
  vol.~103, pp.~2429--2434, Apr. 1999.

\bibitem{colvin_electric_1994}
V.~L. Colvin, K.~L. Cunningham, and A.~P. Alivisatos, ``Electric field
  modulation studies of optical absorption in {CdSe} nanocrystals: Dipolar
  character of the excited state,'' {\em The Journal of Chemical Physics},
  vol.~101, pp.~7122--7138, Oct. 1994.

\bibitem{empedocles_quantum-confined_1997}
S.~A. Empedocles and M.~G. Bawendi, ``Quantum-confined stark effect in single
  {CdSe} nanocrystallite quantum dots,'' {\em Science}, vol.~278,
  pp.~2114--2117, Dec. 1997.

\bibitem{wrigge_efficient_2008}
G.~Wrigge, I.~Gerhardt, J.~Hwang, G.~Zumofen, and V.~Sandoghdar, ``Efficient
  coupling of photons to a single molecule and the observation of its resonance
  fluorescence,'' {\em Nature Physics}, vol.~4, pp.~60--66, Jan. 2008.

\bibitem{devoret_amplifying_2000}
M.~H. Devoret and R.~J. Schoelkopf, ``Amplifying quantum signals with the
  single-electron transistor,'' {\em Nature}, vol.~406, pp.~1039--1046, Aug.
  2000.

\bibitem{kinkhabwala_large_2009}
A.~Kinkhabwala, Z.~Yu, S.~Fan, Y.~Avlasevich, K.~M\"{o}llen, and W.~E. Moerner,
  ``Large single-molecule fluorescence enhancements produced by a bowtie
  nanoantenna,'' {\em Nature Photonics}, vol.~3, pp.~654--657, Nov. 2009.

\bibitem{betzig_proposed_1995}
E.~Betzig, ``Proposed method for molecular optical imaging,'' {\em Optics
  Letters}, vol.~20, pp.~237--239, Feb. 1995.

\bibitem{hettich_nanometer_2002}
C.~Hettich, C.~Schmitt, J.~Zitzmann, S.~K\"{u}hn, I.~Gerhardt, and
  V.~Sandoghdar, ``Nanometer resolution and coherent optical dipole coupling of
  two individual molecules,'' {\em Science}, vol.~298, pp.~385--389, Oct. 2002.

\bibitem{weisenburger_cryogenic_2014}
S.~Weisenburger, B.~Jing, D.~H\"{a}nni, L.~Reymond, B.~Schuler, A.~Renn, and
  V.~Sandoghdar, ``Cryogenic colocalization microscopy for nanometer-distance
  measurements,'' {\em {ChemPhysChem}}, vol.~15, pp.~763--770, Mar. 2014.

\bibitem{middleton_collective_1993}
A.~A. Middleton and N.~S. Wingreen, ``Collective transport in arrays of small
  metallic dots,'' {\em Physical Review Letters}, vol.~71, pp.~3198--3201, Nov.
  1993.

\bibitem{gorter_possible_1951}
C.~J. Gorter, ``A possible explanation of the increase of the electrical
  resistance of thin metal films at low temperatures and small field
  strengths,'' {\em Physica}, vol.~17, pp.~777--780, Aug. 1951.

\bibitem{dayen_enhancing_2013}
J.-F. Dayen, E.~Devid, M.~V. Kamalakar, D.~Golubev, C.~Gu\'{e}don,
  V.~Faramarzi, B.~Doudin, and S.~J. van~der Molen, ``Enhancing the molecular
  signature in molecule-nanoparticle networks via inelastic cotunneling,'' {\em
  Advanced Materials}, vol.~25, no.~3, p.~400–404, 2013.

\bibitem{sadewasser_kelvin_2011}
S.~Sadewasser and T.~Glatzel, {\em Kelvin Probe Force Microscopy: Measuring and
  Compensating Electrostatic Forces}.
\newblock Springer, Oct. 2011.

\bibitem{levitov_electron_1996}
L.~S. Levitov, H.~Lee, and G.~B. Lesovik, ``Electron counting statistics and
  coherent states of electric current,'' {\em Journal of Mathematical Physics},
  vol.~37, pp.~4845--4866, Oct. 1996.

\bibitem{bylander_current_2005}
J.~Bylander, T.~Duty, and P.~Delsing, ``Current measurement by real-time
  counting of single electrons,'' {\em Nature}, vol.~434, pp.~361--364, Mar.
  2005.

\bibitem{nicolet_intermolecular_2006}
A.~Nicolet, M.~A. Kol’chenko, B.~Kozankiewicz, and M.~Orrit, ``Intermolecular
  intersystem crossing in single-molecule spectroscopy: Terrylene in anthracene
  crystal,'' {\em The Journal of Chemical Physics}, vol.~124, p.~164711, Apr.
  2006.

\end{thebibliography}

\end{document}